\documentclass[11pt]{article}
\usepackage{epsfig,times} 
\usepackage{picinpar}
\usepackage{wrapfig}
\usepackage{floatflt}
\makeatletter
\renewcommand{\section}{\@startsection{section}{1}{0in}
	{0.4\baselineskip}{0.1\baselineskip}{\Large\bf}}
\renewcommand{\subsection}{\@startsection{subsection}{2}{0in}
	{0.25\baselineskip}{-\baselineskip}{\large\bf}}
\renewcommand{\subsubsection}{\@startsection{subsubsection}{3}{0in}
	{0.1\baselineskip}{-\baselineskip}{\normalsize\bf}}
\makeatother
\begin{document}

%
\thispagestyle{myheadings}
%
\markright{HE 6.3.14}
\begin{center}
%
{\LARGE \bf Radiodetection of Neutrino Interactions in Ice.}
\end{center}

\begin{center}
%
%
{\bf J. Alvarez-Mu\~niz, R.A. V\'azquez, and E. Zas}\\
{\it Dept. F\'\i sica de Part\'\i culas, Universidade de Santiago, E-15706 
Santiago de Compostela, SPAIN}
\end{center}

\begin{center}
{\large \bf Abstract\\}
\end{center}
\vspace{-0.5ex}
We study the Cherenkov radio pulses emitted in PeV and EeV neutrino
interactions in ice. We discuss how the rich radiation pattern in the
100 MHz to 1 GHz frequency range, in principle, allows the measurement of
shower elongation produced in neutrino interactions opening up the
possibility of flavor recognition through the identification of  
charged current electron neutrino interactions. 
\vspace{1ex}

%
%
\section{Introduction}
\label{intro.sec}

High energy neutrino detection is one of the most attractive challenges for 
the next millennium. EeV neutrino fluxes are unavoidable in the production of 
the highest energy cosmic rays and their interactions with the Cosmic Microwave
Background. There are several experimental efforts to construct Cherenkov
detectors under water or ice with typical sizes of the order of $1~{\rm km^3}$.
The detection of high energy neutrinos could test models of acceleration and
propagation of the highest energy cosmic rays. Moreover it can  
probe unexplored regions in the parameter space for neutrino
oscillations. The expected neutrino fluxes are however low and could need 
very large installations. In this context it is worth exploring alternative 
techniques such as horizontal showers or the detection of Cherenkov radio 
pulses from neutrino interactions which could definitely provide complementary 
information to muon underground detectors.  

The radio technique was proposed in the late fifties (Askar'yan 1962) as a 
possible means for detecting high energy showers. When the wavelength of the 
emitted radiation is greater than the physical shower dimensions, 
the emission from all particles is coherent and the electric field
becomes proportional to the excess charge and hence to shower energy.  
The power radiated in radio thus scales with the square of the shower 
energy making the technique most attractive for detecting showers of 
the highest energies. Radio pulses have been observed in air showers 
but systematic studies are prevented by the influence of distant 
atmospheric phenomena which are difficult to control. This is not expected 
to be a problem for the detection of high energy showers produced by 
neutrinos in denser media because the showers are reduced by the density  
fraction and charge excess in the shower is then expected to be the 
dominant production mechanism for radio pulses. 
Ice has been proposed as an appropriate medium because of its large 
attenuation length for radio signals ($\sim$ 1 km) (Markov 1986, Ralston 1989)
and efforts are being made to test its viability in Antarctica (Allen 1997). 
Among the potential advantages of the technique are the relatively 
low cost of the detectors (antennae) and most importantly the fact that 
information about the development of the excess charge in the shower can 
in principle be recovered from the rich diffraction pattern produced. 
In this work we calculate the different patterns produced by charged 
current electron neutrino interactions compared to the rest as a way 
to illustrate the potential of searching for coherent radio pulses. We 
briefly address the strategies to obtain information from an array of 
antennas. 

\section{Cherenkov radiation from neutrino showers}
\label{cherenkov.sec}

When an electron neutrino interacts with a nucleus through a charged current
deep inelastic interaction (DIS), it produces an electron which initiates an 
electromagnetic shower while simultaneously a hadronic shower is initiated 
by the nuclear fragments. While these interactions produced ''mixed'' 
showers, in neutral current interactions of any flavor neutrinos and 
in charged current interactions induced by muon neutrinos only the hadronic 
shower is expected. We have previously shown that at high energies the 
Landau-Pomeranchuk-Migdal (LPM) effect affects electromagnetic (Alvarez 1997) 
and hadronic (Alvarez 1998, Alvarez 1999) showers in quite different ways.    

The LPM manifests as a dramatic reduction of the pair production and 
bremsstrahlung cross sections at energies above $E_{\rm LPM}$ ($2$~PeV in ice) 
due to large scale correlations in the atomic electric fields. It 
dramatically elongates the development of electromagnetic showers in ice 
for energies above 20 PeV (Alvarez 1997). This can be quantified by an 
increase  in shower length \footnote{Defined as the length along which shower 
size exceeds $70\%$ of its maximum.} proportional to $E_0^{1/3}$, where $E_0$ 
is shower energy. EeV hadronic showers initiated in neutrino interactions 
are less affected by the LPM effect (Alvarez 1998, Alvarez 1999). 
Below 1 EeV they do not display the typical elongation due to the LPM, while 
only a fraction of showers with energy above 1 EeV, show LPM tails which 
typically contain $10\%$ of the shower energy. 
\begin{figwindow}[1,r,%
{\mbox{\epsfig{file=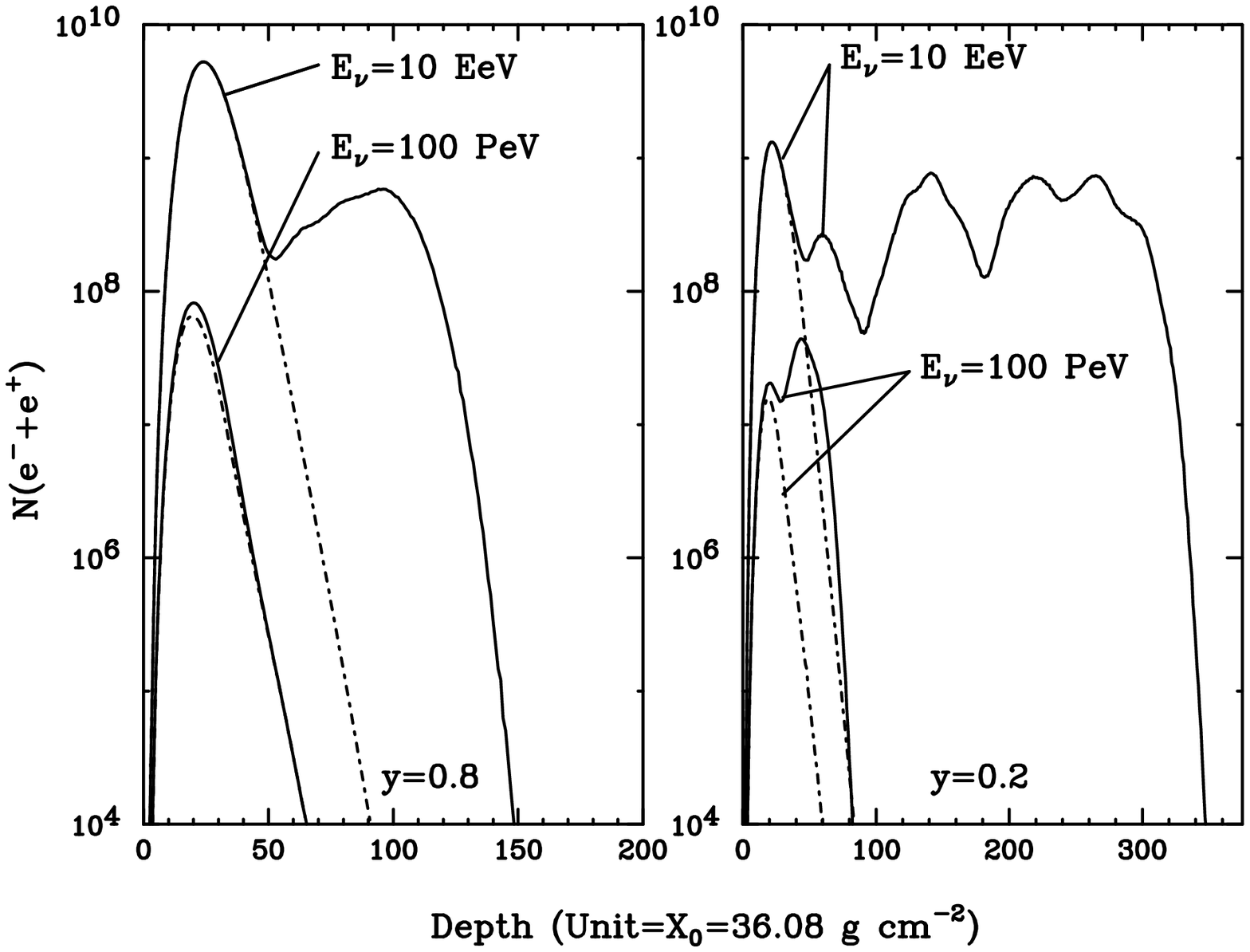,width=5.2in}}},%
{Longitudinal development of hadronic (dashed lines) and mixed showers 
(continuous lines) initiated by neutrino interactions in ice for different 
values of neutrino energies and $y$, as marked.}]
The mixed character showers due to charge current electron interactions 
display the typical features due to the LPM effect as long
as the electromagnetic component has an energy greater than 20 PeV. This 
depends on neutrino energy and on the energy transfer to the hadron debris.
In Figs. 1 and 2 we show the longitudinal development of hadronic and 
mixed showers for neutrino energies 100 PeV and 10 EeV, and fractions of
energy $y=0.2$ and $y=0.8$ respectively. The differences between the mixed and 
hadronic showers become evident as both the neutrino energy and the fraction 
of energy transferred to the electromagnetic component increase.  
\end{figwindow}

All the charged particles in the shower that travel at a speed greater
than the speed of light in the medium contribute to the emission of 
Cherenkov radiation. When the wavelength is in the optical regime, the 
radiation is emitted incoherently. The output power is proportional to the 
total tracklength and hence to shower energy but in the radio regime the 
electric field amplitude is proportional to the excess charge in the shower 
because of coherence. In this case the electric field spectrum at a fixed 
frequency considered as a function observation direction exhibits a 
diffraction pattern which resembles that of a slit with a maximum in the 
Cherenkov direction ($\theta_C=56^{\rm o}$ in ice). The width of the 
Cherenkov peak is inversely proportional to shower length and to the 
frequency. 

For a given observation angle the frequency spectrum of the electric field  
rises linearly with frequency up to a maximum frequency at which coherence 
is lost. In the Cherenkov direction this maximum frequency 
is mostly due to the lateral structure of the shower while away from it, 
it is due to interference between different stages in shower development, 
i.e. to the longitudinal shower development. It can be shown that spectrum of 
the electric field amplitude close to the Cherenkov direction can be well 
approximated by the Fourier transform of the 1-Dimensional longitudinal 
development of the charge excess. We have made
extensive checks to confirm the validity of this approximation, comparing its
results with those obtained with full simulations. 
This method has allowed the study of the angular distributions of the radio 
pulse spectrum using fast programs to calculate shower development for 
energies up to 100 EeV (Alvarez 1997, Alvarez 1998, Alvarez 1999). 

\begin{figwindow}[1,r,%
{\mbox{\epsfig{file=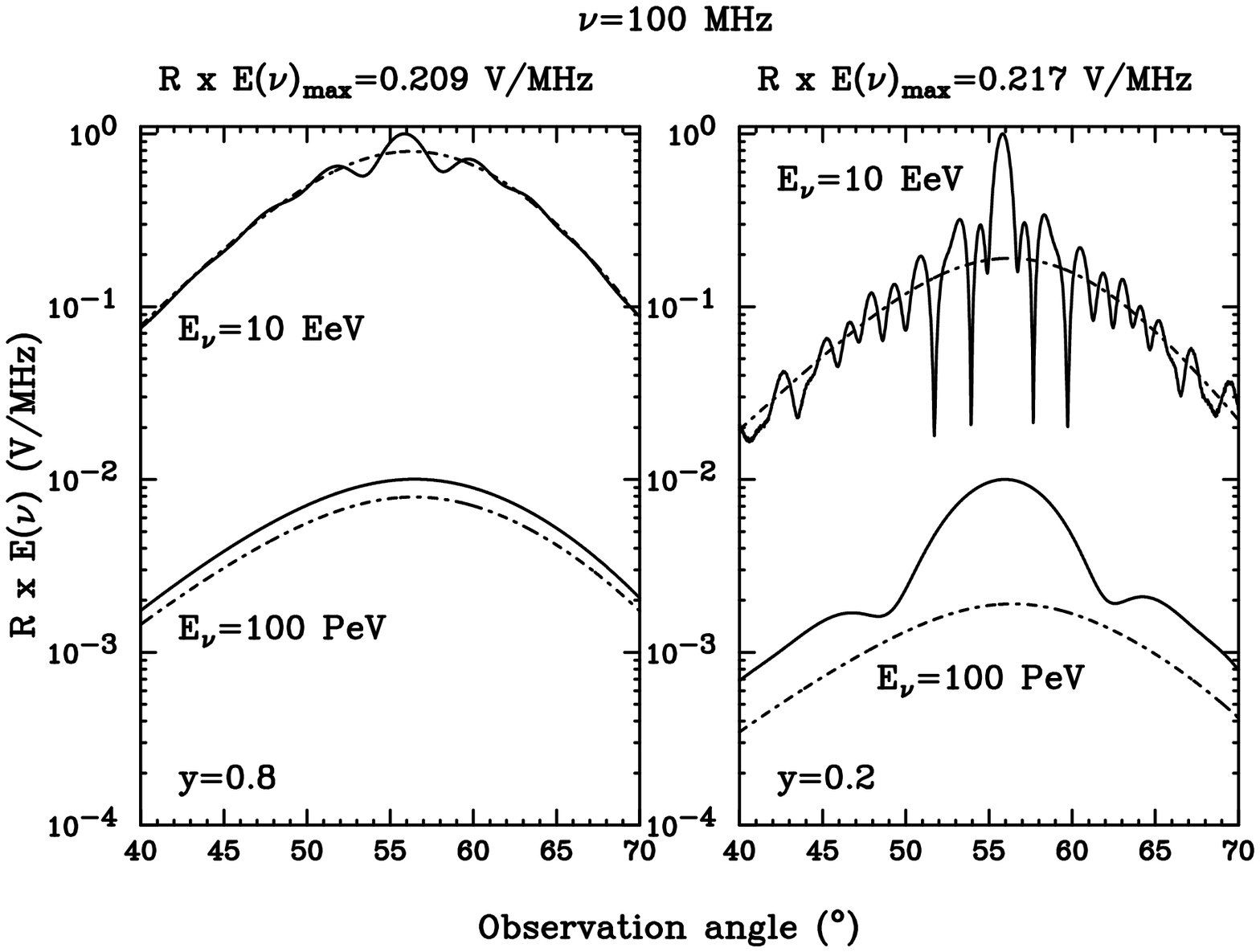,width=5.2in}}},%
{Electric field amplitude around the Cherenkov angle corresponding to 
the showers of Figure 1.}]
In Fig.~2 we show the angular distribution of the Fourier transform of the 
electric field emitted by the same showers in Fig.~1. The electric field 
normalization is chosen to be one in the Cherenkov direction for the shower 
with highest electromagnetic content so that relative amplitudes are correctly 
plotted. As shower length only increases by $\sim~30\%$ between 100 TeV and 10
EeV hadronic showers, the angular distribution narrows approximately in the 
same proportion, a hardly noticeable effect. For the mixed showers induced by 
charged current electron neutrino interactions the situation is quite 
different. If the electromagnetic component has an energy $E_{em}$ exceeding 
20 PeV, the peak narrows as $E_{em}^{1/3}$ (Alvarez 1999). As the average 
fraction of energy transferred to the nuclear fragments is expected to be 
about $<y>=0.25$, the LPM effect due to the electromagnetic part will be 
typically apparent already for electron neutrinos of energy above 
$\sim 30~$PeV. These showers display a characteristic diffraction pattern from 
the superposition of narrow and wide patterns due to the electromagnetic and 
hadronic components respectively. 
\end{figwindow}

\section{The potential of the radio technique}

The radio technique has a large potential for high energy neutrino astronomy 
because of the rich structure of the electric field frequency spectrum and its 
angular diffraction. Information on neutrino interactions can be obtained with 
an array of antennas covering a large region and possibly using 
radiotelescopes pointing to the Moon (Dagkesamansky 1991, Alvarez 1997b, 
Gorham 1999).

The relative arrival times of the radio signals to four antennas allows 
the reconstruction of the radio emission source. 
Once the shower has been located the amplitudes measured in different 
detectors can be used to reconstruct the angular behavior of the electric 
field amplitude from which the shower direction is established 
ideally through the recognition of the direction of highest radiation, 
that is the region ''illuminated'' by the Cherenkov peak. 
Moreover the electric field amplitude measurement in the Cherenkov 
direction is proportional to the tracklength due to the excess charge 
in the shower and hence to the total electromagnetic energy in the shower. 
As distance to the shower is known the measurement allows an accurate 
determination of electromagnetic energy in the shower. If the width of the 
Cherenkov peak can be reconstructed then shower length should be established 
indicating whether there has been a significant elongation due to the LPM 
that is establishing whether high energy photons or electrons have been 
observed. This information can be converted to electron neutrino recognition 
provided the neutrinos have energies in excess of 30~PeV 

There are however more ways to extract useful information from the 
interaction which can be considered as redundant but can be of extreme value 
for reducing spurious signals from other sources, what has been argued can be 
a serious problem for the technique. 
A suitable array of antennas should establish the Cherenkov cone region 
and from its geometrical shape both the shower position and orientation 
should become readily available.  
Cherenkov radiation is polarized in the 
direction of apparent movement of the excess charge and this can be very 
valuable (Jelley 1996). If the polarization of the signal is measured at 
three separate locations the emission source can also be located. 
The technique even allows obtaining a fair amount of information from a 
single site. The frequency spectrum of the electric field 
increases linearly with frequency until a maximum value which depends on the 
observation angle. Sampling the frequency spectrum at a single location 
allows the determination of the angle between the observation direction 
and the Cherenkov directions provided the shower is of hadronic type and 
in any case the amplitude measurements corresponding to the linear region 
of the spectrum also allow the determination of electromagnetic shower energy 
regardless of the type of shower in question.  

In Figs. 1 and 2 we display the shower development and the corresponding 
angular distribution of the radio pulse spectrum. The figures illustrate 
the potential of the radio technique, and particularly how 
$\nu_e$ charged current interactions can be distinguished  
from neutral current interactions or 
charge current interactions of muon neutrinos. 
This is a characteristic of the radio technique due to its coherent 
character so that a wealth of information is carried by the rich diffraction 
pattern generated. In principle a simple  
inverse Fourier transform should allow the reconstruction 
of much longitudinal shower development what could be used to distinguish 
electromagnetic, hadronic and mixed type showers and hence 
opening up the possibility of flavor recognition.  
Surely the possibility would put much more stringent constraints on the 
design, rising its cost an complexity by large factors and this would have to 
wait till the technique is shown to be viable, however the wealth of 
information that could be obtained from such technique surely makes the 
effort worth its while.  

We thank F.~Halzen for suggestions after reading the 
manuscript and G.~Parente, T. Stanev, and I.M. Zheleznykh for helpful 
discussions. This work was supported in part by CICYT (AEN96-1773) and by 
Xunta de Galicia (XUGA-20604A96). J.A. thanks the Xunta de Galicia for 
financial support.

%
\vspace{1ex}
\begin{center}
{\Large\bf References}
\end{center}
Allen C. {\it et al.} astro-ph/9709223. \\
Alvarez-Mu\~niz, J. and \& Zas, E, 1999 These proceedings (HE 2.5.31)\\
Alvarez-Mu\~niz, J., \& Zas, E., 1997, Phys. Lett. B 411, p.218\\
Alvarez-Mu\~niz, J., \& Zas, E., 1998, Phys. Lett. B 411, p.396\\
Askar'yan,~G.A., {\sl Soviet Physics} JETP {\bf 14,2} 441
 (1962); {\bf 48 } 988 (1965).\\
Dagkesamansky, R.D. and Zeleznykh,~I.M., {\sl Proc. of the ICRR International
Symposium: Astrophysical Aspects of the most energetic Cosmic Rays}, World 
Scient., eds. M. Nagano and F. Takahara, 1991, p.373. \\
Gaisser, T.K., 1990, {\it Cosmic Rays and Particle Physics},
Cambridge University Press\\
Gorham,~P. {\it et al.}, These Proceedings\\
Jelley, J.V., {\sl Astro. Phys.} {\bf 5} 255 (1996). \\
Markov,~M.A. and Zeleznykh,~I.M., {\sl Nucl.\ Inst.\ Methods}
{\bf A 248} 242--251 (1986)\\
Ralston,~J.P. and McKay,~D.M.,  {\sl Proc.\ Astrophysics in
Antarctica Conference}, ed.\ Mullan,~D.J., Pomerantz,~M.A. and
Stanev,~T., (American Institute of Physics, New York, 1989) Vol.~198. p.~241 \\
\end{document}